\documentclass[slac_one]{revtex4}
\usepackage{graphicx}
\usepackage{fancyhdr}
\def\etal{{\it et.al.}}

\begin{document}

\title{{\small{2005 International Linear Collider Workshop - Stanford,
U.S.A.}}\\ 
\vspace{12pt}
Neutrinos in Supersymmetry} 

\author{Marco Aurelio D\'\i az}
\affiliation{Department of Physics, Universidad Cat\'olica de Chile,
Santiago, Chile}

\begin{abstract}
We briefly review the neutrino mass generation mechanism in supersymmetry 
with Bilinear R-Parity Violation in Minimal Supergravity and Anomaly 
Mediated Supersymmetry Breaking.

\end{abstract}

\maketitle

\thispagestyle{fancy}


\section{INTRODUCTION} 

Experimental results indicate that neutrinos oscillate: the three known 
neutrino flavour states $\nu_e$, $\nu_{\mu}$, and 
$\nu_{\tau}$, are linear combinations of the mass eigenstates $\nu_1$,
$\nu_2$, and $\nu_3$, inducing neutrino oscillations in vacuum 
\cite{vacuum} and in matter \cite{matter}. The experimental results 
come from many experiments such as, Kamiokande, SAGE, GALLEX, GNO, 
Super-Kamiokande, and SNO  for solar neutrino \cite{Fukuda:1996sz}; 
KamLAND for long-baseline reactor neutrino  \cite{Eguchi:2002dm}; 
Kamiokande, Super-Kamiokande, MACRO, and Soudan-2 for atmospheric 
neutrino \cite{Hatakeyama:1998ea}; K2K for long-baseline accelerator 
neutrino \cite{Ahn:2002up}; and CHOOZ and Palo Verde for short-baseline 
accelerator neutrino experiments \cite{Apollonio:1999ae}.

The neutrino mass matrix is diagonalized with the matrix 
\cite{Fogli:2005cq}
\begin{equation}
U_{PMNS}= 
\left(\begin{array}{ccc}
  1 &                0 &               0 \\
  0 &  \cos\theta_{23} & \sin\theta_{23} \\
  0 & -\sin\theta_{23} & \cos\theta_{23}
\end{array}\right)
\left(\begin{array}{ccc}
  \cos\theta_{13} & 0 & \sin\theta_{13} \\
                0 & 1 &               0 \\
 -\sin\theta_{13} & 0 & \cos\theta_{13}
\end{array}\right)
\left(\begin{array}{ccc}
  \cos\theta_{23} & \sin\theta_{23} & 0\\
 -\sin\theta_{23} & \cos\theta_{23} & 0\\
                0 &               0 & 1
\end{array}\right)
\end{equation}
This matrix may contain also a Dirac and two Majorana phases 
\cite{Schechter:1980gr}, which we assume to be absent. The oscillations
are defined by the atmospheric $\theta_{23}$, solar $\theta_{12}$ and 
reactor $\theta_{13}$ mixing angles, and the atmospheric 
$\Delta m^2_{32}$ and solar $\Delta m^2_{21}$ mass squared differences.

There are several analysis of these experimental results 
\cite{Nunokawa:2002mq}. We use the $3\sigma$ allowed regions for the 
neutrino parameters in \cite{Maltoni:2004ei}, given by
\begin{eqnarray}
1.4 \times10^{-3} < \Delta m^2_{32} < 
3.3 \times10^{-3}\,{\mathrm{eV}}^2
& \qquad &
0.52 < \tan^2\theta_{23} < 2.1
\nonumber\\
7.2 \times10^{-5} < \Delta m^2_{21} < 
9.1 \times10^{-5}\,{\mathrm{eV}}^2
& \qquad &
0.30 < \tan^2\theta_{12} < 0.61
\label{3s5sLim}
\end{eqnarray}
which we complete with the upper bound $\tan^2\theta_{13} < 0.049$ for
the reactor angle.

\section{LOW ENERGY SEESAW MECHANISM}

We study here the generation of neutrino masses in supersymmetry with
Bilinear R-Parity Violation (BRpV). The superpotential of our model 
differs from the MSSM by three terms which violate R-Parity and lepton 
number \cite{Diaz:1997xc},
\begin{equation}
W=W_{MSSM}+\epsilon_i \hat L_i \hat H_u
\end{equation}
where $\epsilon_i$ have units of mass. These bilinear terms induce
mixing between the neutralinos and neutrinos, forming a $7\times7$ mass 
matrix. A low energy seesaw mechanism induces the following effective 
$3\times3$ neutrino mass matrix
\begin{equation}
{\bf M}_{\nu}^{(0)}=
\frac{M_1 g^2 \!+\! M_2 {g'}^2}{4\, det({\cal M}_{\chi^0})}
\left[\matrix{
\Lambda_1^2 & \Lambda_1 \Lambda_2 & \Lambda_1 \Lambda_3 \cr
 \Lambda_1 \Lambda_2 & \Lambda_2^2 & \Lambda_2 \Lambda_3 \cr
\Lambda_1 \Lambda_3 & \Lambda_2 \Lambda_3 & \Lambda_3^2
}\right]
\end{equation}
where we have defined the parameters $\Lambda_i=\mu v_i+\epsilon_i v_d$,
which are proportional to the sneutrino vacuum expectation values in the 
basis where the $\epsilon$ terms are removed from the superpotential.
At tree-level only one neutrino acquire a mass, and one-loop corrections
must be included \cite{Romao:1999up}. 

\section{ANOMALY MEDIATED SUSY BREAKING}

In AMSB-BRpV we work with \cite{deCampos:2004iy,Diaz:2002ij}
\begin{equation}
  m_{3/2}=35 \,{\mathrm{TeV}}\,,\quad 
  m_0=250 \,{\mathrm{GeV}}\,,\quad 
  \tan\beta=15\,,\quad \hbox{ and} \quad
 {\mathrm{sign}}(\mu) < 0 \; .
\label{refAMSB}
\end{equation}
and we randomly vary the parameters $\epsilon_i$ and $\Lambda_i$ looking
for solutions in which the restrictions (\ref{3s5sLim}) from neutrino 
physics are satisfied. An example of these solutions is
\begin{eqnarray}
  \epsilon_1 = -0.015 \hbox{ GeV}   \;,  \qquad 
& \epsilon_2 = -0.018 \hbox{ GeV}   \;,  \qquad 
& \epsilon_3 =  0.011 \hbox{ GeV}   \;,
\nonumber\\
  \Lambda_1  = -0.03  \hbox{ GeV}^2 \; ,  \qquad 
& \Lambda_2  = -0.09  \hbox{ GeV}^2 \; ,  \qquad 
& \Lambda_3  = -0.09  \hbox{ GeV}^2 \; . 
\label{basic:brpv}
\end{eqnarray}
The neutrino parameters obtained in this reference model are
\begin{eqnarray}
 \Delta m^2_{\mathrm{atm}}=2.4\times10^{-3}\,{\mathrm{eV}}^2\,,\qquad 
& \tan^2\theta_{\mathrm{atm}}=0.72\;, \qquad 
& \tan^2 \theta_{13} = 0.033 \; ,
\nonumber\\
 \Delta m^2_{\mathrm{sol}}=7.9\times10^{-5}\,{\mathrm{eV}}^2\,,\qquad 
&  \tan^2\theta_{\mathrm{sol}}=0.47\;, \qquad 
&
\label{NeuRes}
\end{eqnarray}
which agree with the present experimental results. The neutrino mass 
matrix has the following texture
\begin{equation}
{\bf M}^{eff}_{\nu}=m\left[\matrix{
\lambda  & 2\lambda & \lambda \cr
2\lambda & a        & b       \cr
\lambda  & b        & 1
}\right]
\label{texture1}
\end{equation}
with $a\sim 0.74$, $b\sim 0.67$, $\lambda\sim 0.13$, and $m\sim0.032$ eV. 

\begin{figure}
\centering
\includegraphics[width=80mm]{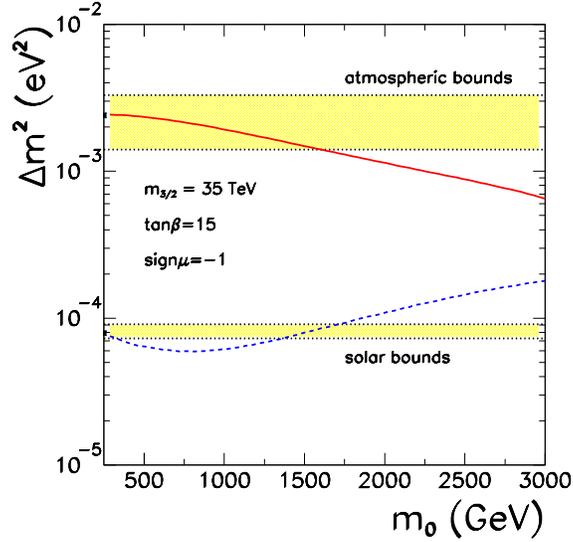}
\caption{
  The red solid (blue hashed) line stands for the predicted atmospheric
  (solar) mass squared difference as a function of the scalar mass $m_0$ 
  for $m_{3/2}=35$ TeV, $\tan\beta=15$, and $\hbox{sign}(\mu) < 0$ and 
  for the BRpV parameters given in (\ref{basic:brpv}). The allowed 
  $3\sigma$ atmospheric (solar) mass squared difference is represented 
  by the upper (lower) horizontal yellow band. Our reference point is 
  represented by a star on the left of the plot.
} \label{m0-1}
\end{figure}
In Fig.~\ref{m0-1} it is shown the dependence on the scalar mass $m_0$ 
of the predicted atmospheric neutrino mass squared difference 
$\Delta m^2_{\mathrm{atm}}$ (red solid line) and the solar neutrino mass 
squared difference $\Delta m^2_{\mathrm{\mathrm{sol}}}$ (blue dashed 
line), for fixed values $m_{3/2}=35$ TeV, $\tan\beta=15$, and $\hbox{sign}
(\mu) < 0$ and for the BRpV parameters given in (\ref{basic:brpv}).  
We see from Fig.~\ref{m0-1} that $\Delta m^2_{\mathrm{atm}}$ is
within the present experimental bounds for $m_0 \lesssim 1.6$ TeV while
$\Delta m^2_{\mathrm{sol}}$ satisfies the experimental constraints for $m_0
\lesssim 310$ GeV and $ 1.4 \hbox{ TeV} \lesssim m_0 \lesssim 1.75$ TeV.
Therefore, our models lead to acceptable neutrino masses provided $m_0
\lesssim 310$ GeV or $ 1.4 \hbox{ TeV} \lesssim m_0 \lesssim 1.6$ TeV for all
other parameters fixed at their reference values.  It is also important to
notice that the heaviest neutrino state has a mass of the order of $0.050$ eV
for our reference point and that it decreases as $m_0$ increases. Moreover,
the radiative corrections lead to a contribution of ${\cal O}(10\%)$,
therefore, the tree--level result for the neutrino mass is a good order of
magnitude estimative.

\begin{figure}
\centering
\includegraphics[width=80mm]{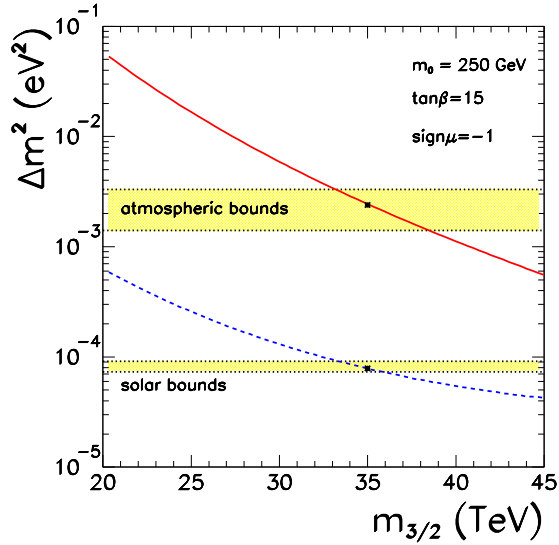}
\caption{
  Atmospheric (solid line) and solar (dashed line) mass squared 
  differences as a function of the gravitino mass $m_{3/2}$. The 
  remaining parameters assume the value of our reference point and the 
  conventions are the same of Fig.~\ref{m0-1}.
} \label{m32-1}
\end{figure}
In Fig.~\ref{m32-1} we display the dependence of the atmospheric and solar
mass squared differences on the gravitino mass $m_{3/2}$ for the other
parameters assuming their reference values. First of all, the observed
dependence is much stronger compared to the dependence on $m_0$; this is
expected due to the large impact of $m_{3/2}$ on the soft gaugino masses,
which together with $\mu$ define the tree--level neutrino mass matrix.
Moreover, the SUSY spectrum has a large impact on the one--loop corrections
increasing the sensitivity to $m_{3/2}$.  Both solar and atmospheric 
squared mass differences are too large in the region of small gravitino 
masses, however, this region is already partially ruled out since it leads 
to charginos lighter than the present experimental bounds for $m_{3/2} 
\lesssim 30$ TeV.  Conversely, there is no acceptable solution for the 
neutrino masses at large $m_{3/2}$, again a region partially ruled out by 
data since the staus are too light in this region.  Furthermore, we can 
see from this figure that our AMSB--BRpV model leads to acceptable neutrino 
masses for a small window of the gravitino mass ($33\; \rm{TeV}\lesssim 
m_{3/2} \lesssim 36\;\rm{TeV}$) given our choice of parameters. This is 
far from trivial since we have no {\em a priori} guaranty that we can 
generate the required neutrino spectrum, specially the radiative 
corrections, satisfying at the same time the experimental constraints on 
the superpartner masses.

\section{MINIMAL SUPERGRAVITY}

Our analysis in Sugra-BRpV is defined by \cite{Diaz:2004fu}
\begin{equation}
m_0=100\,{\mathrm{GeV}} \,,\, M_{1/2}=250\,{\mathrm{GeV}} \,, \,
A_0=-100\,{\mathrm{GeV}} \,,\, \tan\beta=10 \,,\, \mu>0
\label{sugBench}
\end{equation}
where the neutralino is the LSP with a mass $m_{\chi^0_1}=99$ GeV, and 
the light neutral Higgs boson with $m_{h}=114$ GeV.

In this context we find several solutions for neutrino physics which satisfy 
the experimental constraints on the atmospheric and solar mass squared 
differences, and the three mixing angles. We single out the following
\begin{eqnarray}
\epsilon_1=-0.0004 \,,\qquad &\epsilon_2=0.052 \,,\qquad 
&\epsilon_3=0.051 \,,\, 
\nonumber\\
\Lambda_1=0.022 \,,\qquad &\Lambda_2=0.0003 \,,\qquad 
&\Lambda_3=0.039 \,,\, 
\label{epslam}
\end{eqnarray}
This solution is characterized by 
\begin{eqnarray}
 \Delta m^2_{\mathrm{atm}}=2.7\times10^{-3}\,{\mathrm{eV}}^2\,,\qquad 
& \tan^2\theta_{\mathrm{atm}}=0.72\;, \qquad 
& \tan^2 \theta_{13} = 0.0058 \; ,
\nonumber\\
 \Delta m^2_{\mathrm{sol}}=8.1\times10^{-5}\,{\mathrm{eV}}^2\,,\qquad 
&  \tan^2\theta_{\mathrm{sol}}=0.54\;, \qquad 
&
\label{NeuRes2}
\end{eqnarray}
which are well inside the experimentally allowed window in 
eq.~(\ref{3s5sLim}). We note that the random solution in 
eq.~(\ref{epslam}) is compatible with $\epsilon_1=\Lambda_2=0$, 
{\it i.e.}, the neutrino parameters in eq.~(\ref{NeuRes2}) are hardly 
changed with this replacement. The neutrino mass matrix has the 
following texture
\begin{equation}
{\bf M}^{eff}_{\nu}=m\left[\matrix{
\lambda & 0 & \lambda \cr
0       & a & a       \cr
\lambda & a & 1
}\right]
\label{texture2}
\end{equation}
with $a\sim 0.75$, $\lambda\sim 0.14$, and $m\sim0.033$ eV. 

\begin{figure}
\centering
\includegraphics[width=80mm,angle=90]{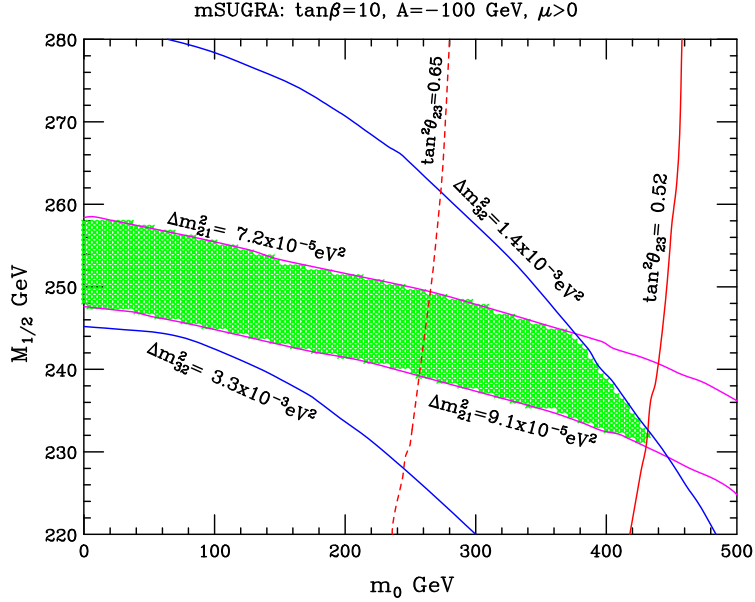}
\caption{
  Region of parameter space in the plane $m_0-M_{1/2}$ where
  solutions to neutrino physics passing all the implemented experimental 
  cuts are located. Contours of constant atmospheric mass difference and 
  angle, and solar mass difference are displayed.
} \label{m0mhalf}
\end{figure}
In Fig.~\ref{m0mhalf} we take the neutrino solution given by the 
BRpV parameters in eq.~(\ref{epslam}), and vary the scalar mass $m_0$ and
the gaugino mass $M_{1/2}$, looking for solutions that satisfy all 
experimental cuts. In this case, sugra points satisfying the 
experimental restrictions on the neutrino parameters lie in the shaded
region. Solutions are concentrated in a narrow band defined by 
$M_{1/2}\approx230-260$ GeV and $m_0\approx0-400$ GeV.
We note that in BRpV the LSP need not to be the lightest neutralino,
since it is not stable anyway. For this reason, the region close to 
$m_0\approx0$ is not ruled out. 

Smaller values of $M_{1/2}$ are not possible because the atmospheric and 
solar mass differences become too large. The allowed strip is, thus, 
limited from below by the curve $\Delta m^2_{21}=9.1\times10^{-5}$ eV${}^2$. 
The dependency on $M_{1/2}$ is felt stronger by the tree level 
contribution. 
Higher values of the scalar mass $m_0$ are not allowed because the 
atmospheric angle becomes too small. The allowed strip is, therefore,
limited from the right by the contour $\tan^2\theta_{23}=0.52$. 
High values of the scalar mass are also limited from above because the 
atmospheric mass becomes too large. 
Higher values of $M_{1/2}$ are not possible because the solar mass becomes
too small, therefore, the allowed stripe is limited from above by the line
$\Delta m^2_{21}=7.2\times10^{-5}$ eV${}^2$. 

\begin{figure}
\centering
\includegraphics[width=80mm,angle=90]{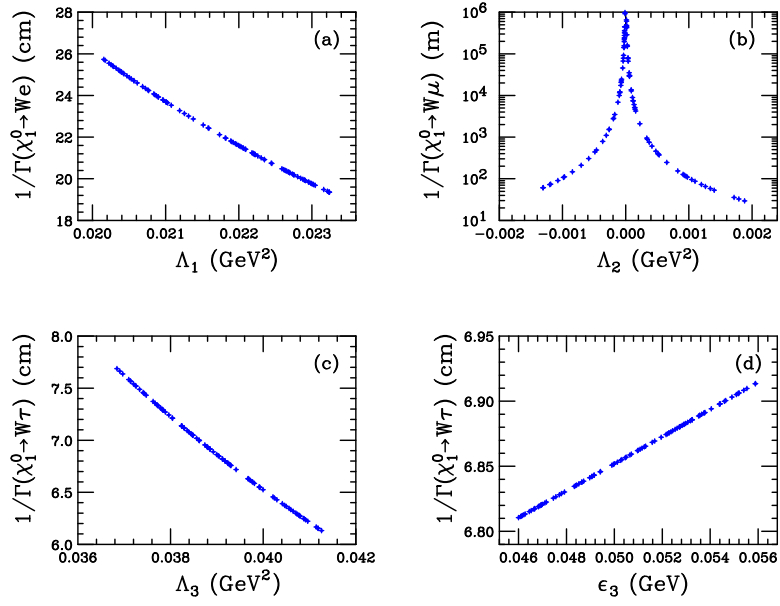}
\caption{
  Partial decay width of a neutralino into a $W$ and a lepton,
  measured in units of distance.
} \label{X1_1}
\end{figure}
In Fig.~\ref{X1_1} we plot the inverse of the partial decay width
(multiplied by the velocity of light to convert it into a distance) as
a function of the most relevant BRpV parameters. In frame (\ref{X1_1}a)
we see the inverse of $\Gamma(\chi^0_1\rightarrow We)$ as a function
of $\Lambda_1$. In fact, for all practical purposes, the decay rate 
into electrons depends {\it only} on $\Lambda_1$. Since in first 
approximation, the coupling is proportional to $\Lambda_1$, the inverse
of the decay rate behaves like $\Lambda_1^{-2}$, and this is seen in the 
figure. The values of $\Lambda_1$ are limited by the solar parameters.
The inverse of the partial decay rate $\chi^0_1\rightarrow We$ is of the 
order of $20-25\, cm$, and it's an important part of the total decay rate.

In frame (\ref{X1_1}b) we have the inverse of 
$\Gamma(\chi^0_1\rightarrow W\mu)$ as a function of $\Lambda_2$, and 
similarly to the previous case, the decay rate into muons depends 
practically only on $\Lambda_2$. In our reference model in 
eq.~(\ref{epslam}) we have $\Lambda_2\approx0$, but values indicated
in the figure are also compatible with neutrino physics. The coupling of 
the neutralino to $W$ and muon is proportional to $\Lambda_2$, so the
inverse of the decay rate goes like $\Lambda_2^{-2}$, and that is 
observed in frame (\ref{X1_1}b). Depending on the value of $\Lambda_2$,
the partial decay length vary from centimeters to kilometers 
in the figure. Therefore, this partial decay rate contribute little 
to the total decay rate of the neutralino.

The inverse of $\Gamma(\chi^0_1\rightarrow W\tau)$ is plotted in frames
(\ref{X1_1}c) and (\ref{X1_1}d) as a function of $\Lambda_3$ and
$\epsilon_3$ respectively. The dependence on $\Lambda_3$ is stronger
and similarly to the previous cases it goes like $\Lambda_3^{-2}$. The 
dependence on $\epsilon_3$ is weaker, and the inverse decay rate 
increases with this parameter. The inverse decay rate is of the order of 
$7\, cm$, making it the most important contribution to the total decay 
rate. Including the decay modes into neutrinos and a $Z$, the total 
inverse decay rate is near $4\, cm$. The ratios of branching ratios for 
our benchmark point in eq.~(\ref{epslam}) are given by 
\begin{equation}
\frac{B(\chi^0_1\rightarrow W\mu)}{B(\chi^0_1\rightarrow W\tau)}=
5.9\times10^{-5}\,,\qquad 
\frac{B(\chi^0_1\rightarrow W e)}{B(\chi^0_1\rightarrow W\tau)}=0.32
\end{equation}
We note that if we increase $\Lambda_2$ by a factor 4, the first ratio of
branching ratios increase to $\sim10^{-3}$ without changing the other 
ratio, while still passing all the experimental cuts. From this figure,
it is clear that by measuring the branching ratios of the neutralinos we 
get information on the parameters of the model.

We calculate the production cross sections 
$\sigma(pp\rightarrow \chi^0_1 \chi^0_1)$ (LHC) and 
$\sigma(e^+ e^-\rightarrow \chi^0_1 \chi^0_1)$ (ILC at $\sqrt{s}=500$ 
GeV) at leading order, and using the branching ratios find
\begin{eqnarray}
  \label{eq:completecrosssections}
  \sigma(pp\rightarrow \chi^0_1 \chi^0_1\rightarrow W^+ W^+ e^-
 \tau^-) &=&3.4\times10^{-4} \mbox{ pb}
\nonumber\\ 
 \sigma(e^+ e^-\rightarrow \chi^0_1 \chi^0_1\rightarrow W^+ W^+ e^-
 \tau^-)&=&9.3\times 10^{-3}\mbox{ pb}
\end{eqnarray}
with negligible background. Assuming a luminosity of 
$10^5$ pb${}^{-1}$/year at both machines, we expect 
$\sim280$ signal event per year at the LHC and $\sim3700$ signal events 
per year at the ILC (summing over lepton charges).

\newpage

\section{CONCLUSIONS}

Supersymmetry with Bilinear R-Parity Violation remains a viable mechanism 
for the generation of neutrino masses and mixing angles. In this model, 
the neutralino is no longer a candidate to dark matter. Nevertheless, it
is possible to understand the neutrino mass spectrum. A low energy seesaw 
mechanism gives mass to one neutrino due to the mixing with neutralinos,
and quantum corrections give mass to the other two. We show examples on 
how this works in supergravity and AMSB, indicating how this model can be 
tested in future colliders.

\begin{acknowledgments}
This research was partly founded by CONICYT grant No.~1040384.
\end{acknowledgments}


\end{document}